# A proof-of-concept online metadata catalogue service of Earth observation datasets for human health research in exposomics

Keumseok Koh [1], Maged N. Kamel Boulos [2,*], Gang Zheng[1], Hongsheng Zhang[1] , Muralikrishna V. Iyyanki[3], Bosco Bwambale[4], Ashraf Dewan[5]

[1]   Dept. of Geography, Faculty of Social Sciences, The University of Hong Kong, Hong Kong 999077, China
[2]   School of Medicine, University of Lisbon, 1649-028 Lisbon, Portugal
[3]   JN Technological University Hyderabad, Hyderabad, 500008, India
[4]   Mountains of The Moon University, P. O. Box 837, Fort-Portal, Uganda
[5]   School of Earth & Planetary Sciences, Curtin University, Perth, WA 6845, Australia
*   Correspondence (M.N.K.B.): mnkboulos@ieee.org

**Abstract:** This article describes research carried out during 2023 under an International Society for Photogrammetry and Remote Sensing (ISPRS)-funded project to develop and disseminate a metadata catalogue of Earth observation data sources/products and types that are relevant to human health research in exposomics, as a free service to interested researchers worldwide. The proof-of-concept catalogue was informed by input from existing research literature on the subject (desk research), as well as online communications with, and relevant research publications collected from, a small panel (n = 5) of select experts from the academia in three countries (China, UK and USA). It has 90 metadata records of relevant Earth observation datasets (n = 40) and associated health-focused research publications (n = 50). The project's online portal offers a searchable version of the catalogue featuring a number of search modes and filtering options. It is hoped future, more comprehensive versions of this service will enable more researchers and studies to discover and use remote sensing data about population-level exposures to disease determinants (exposomic determinants of disease) in combination with other relevant data to reveal fresh insights that could improve our understanding of relevant diseases, and hence contribute to the development of better-optimized prevention and management plans to tackle them.

## 1. Introduction

*1.1. Earth Observation Data*

Earth observation data include remote sensing (RS) data from satellites, aerial platforms, drones, and seaborne and ground-based sensors, among others. Typically obtained by capturing and analyzing the energy reflected or emitted from the Earth's surface, atmosphere, or other objects of interest [1], RS data provide valuable insights into various aspects of the Earth's environment, such as air pollutants, land cover, green space, topography, nocturnal outdoor light pollution, and noise pollution, among others. Moreover, advances in wearable sensors and mobile technologies are further expanding the applications of RS. RS data, combined with other human population data collected at multiple scales are being successfully applied to various real-life problems and challenges to better manage them [2].

*1.2. Human Exposomics*

Predisposition and development of various diseases, communicable, such as malaria and dengue fever, and non-communicable, such as diabetes, cardiovascular diseases, various types of cancer and mental health problems, involve a complex interplay between genetic factors (the genome) and environmental and lifestyle parameters that populations are exposed to (the exposome) [3]. When combined with other relevant data, RS data from various sources/products and of different types can help us better map and investigate the latter (exposomic determinants of disease/population-level exposures) [4-8]. However, these Earth observation data sources and types remain a largely untapped resource for many researchers in the fields of population health and medicine, who are not familiar with the potential and value of these data and the unique insights that can be revealed by using them.

This article describes the authors' work on an International Society for Photogrammetry and Remote Sensing (ISPRS)-funded project entitled 'geospatial database for exposomics (development of an online portal and metadata catalogue of Earth observation data types, sources and products for human health research in exposomics)'. This project is funded under ISPRS Scientific Initiatives 2023 (SI2023 Awards) [9], and involves proof concepting and disseminating a much-needed metadata catalogue of Earth observation data sources/products and types that are relevant to human health research as a free service to interested researchers worldwide. Even though the various uses of RS data in large population health studies are not a new thing as such (existing studies actually form the basis of our project), our rationale is to further highlight and facilitate such uses and open them up to many more researchers and studies in the fields of population health and medicine.

**2. Materials and Methods**

*2.1. Catalogue Curation: External Expert Involvement and Metadata Elements*

The project's proof-of-concept catalogue was informed by a rather extensive input from existing research literature on the subject (desk research) as well as online communications with, and relevant research publications collected from, a small panel (n = 5) of select 'stakeholder representatives' from the academia (backgrounds in health/human geography and/or geomatics) in three countries (China, UK and USA) to guarantee some basic form of user engagement and involvement in the project from its very early stages, which is always highly desirable.

This small panel of five academics was a convenience sample, given the project team's human resource, time and budget constraints and the fact that in this project we are only proof concepting the catalogue and it was never our goal to comprehensively catalogue every RS dataset and application in health under the sun. Ten academics were initially invited via email, but only five of them responded. All invited academics were external experts (i.e., not members of the project team) but well-known to us for their leading research into RS applications in health.

Online communications with panel members took place between February and June 2023. The five stakeholder representatives provided us with valuable insights and intelligence about relevant remote sensing applications and available products in public health, in addition to important feedback on our catalogue during its development over two iterations in 2023, which helped us improve and refine it to best serve their needs and those of the wider research community they represent.

The research team compiled the catalogue based on all collected input in the form of an Excel workbook [10]. The catalogue currently has 90 metadata records of relevant Earth observation datasets (n = 40) and associated health-focused research publications (n = 50). Dataset and research publication records are stored in two separate but fully cross-referenced Excel worksheets/database tables. For each catalogued record, the following metadata elements were compiled: dataset name, provider, available year(s)/update frequency, cost (licence), covered geographical areas, resolution used, URL, related research publications, and the health conditions/application areas that featured in those publications.

*2.2. Web Conversion*

Our Web development team then converted the Excel workbook into an online database and developed the project's online portal in MySQL, an open-source relational database management system; Python, a high-level, general-purpose programming language; and Django, an open source, Python-based Web framework. The portal's public user interface was designed with both desktop and mobile devices in mind. The portal also features an access-protected 'Catalogue Management Interface' where the catalogue curator can modify existing catalogue records or add new ones as necessary.

**3. Results**

*3.1. Catalogue Summary Statistics: By Publications (n = 50)*

3.1.1. Journal Categories

Approximately 40% (n = 21) of curated articles (health-focused research publications) in our proof-of-concept catalogue were published in geography-themed journals, followed by public health (n = 13), environment (n = 9), and science-themed (n = 7) journals.

### 3.1.2. Study Areas

The majority of curated studies in our proof-of-concept catalogue are global studies (n = 30). The remaining studies cover North America/USA (n = 9), Europe/UK (n = 7), Asia (n = 2), Africa (n = 1), and the Global Coastal Zone (n = 1). Readers can refer to [10] for detailed information about the curated publications and their corresponding study areas.

### 3.1.3. Study Topics

Table 1 summarizes the research topics of the curated publications in the proof-of-concept catalogue. We broadly classified the topics into three sub-categories or themes—human activities, public health, and the environment. Almost half of the publications (n = 24) investigated various themes of human activities, followed by public health (n = 14), and the environment (n = 12). Readers can refer to [10] for detailed information about which curated publication(s) is/are covering which topic, with full citation details.

**Table 1.** Study topics.

| Application Areas | Study Topics | Number of Works |
| --- | --- | --- |
| Human Activity (including anthropogenic activity affecting the environment) | Children Physical Activity (n = 4), Community-Level Livability (n = 1), Humanitarian Emergencies (n = 1), Land Use Mix Index (n = 1), Landscape Disturbance (n = 1), Nighttime Lights (n = 2), Population Exposure (n = 1), Residential Environments Greenness (n = 1), Transportation Network (n = 1), Urban Expansion (n = 2), Urban Foraging (n = 1), Urban Green Spaces (n = 2), Urban Sustainability (n = 2), Urban Temperatures/Thermal Environment (n = 3), Urban-associated Salmonellae (n = 1) | 24 |
| Public Health | Anxiety Symptoms during the COVID-19 Pandemic (n = 1), Haemorrhagic Fever (n = 4), Hand, Foot, and Mouth Disease (HFMD) Outbreak (n = 1), Human Health and well-being (n = 1), Human Rabies (n = 1), Influenza Virus (n = 1), Malaria Prevention (n = 1), Rural Mental Health (n = 1), Salmonella Serovars and Strains (n = 1), Sociodemographic Environments and Obesity (n = 1), West Nile Virus (n = 1) | 14 |
| Environment | Air Pollution (n = 2), Grassland (Prairie) Ecosystems (n = 2), Mixed Grasslands (n = 1), Mountain Green Cover Index (n = 1), Ocean Harmful Algal Blooms (n = 4), Rural Water (n = 2) | 12 |
| Total | | 50 |

### 3.1.4. Study Publication Years

Half of the articles (n = 25) were published before 2018. Remarkably, 25% of the articles (n = 12) were published in 2020 alone.

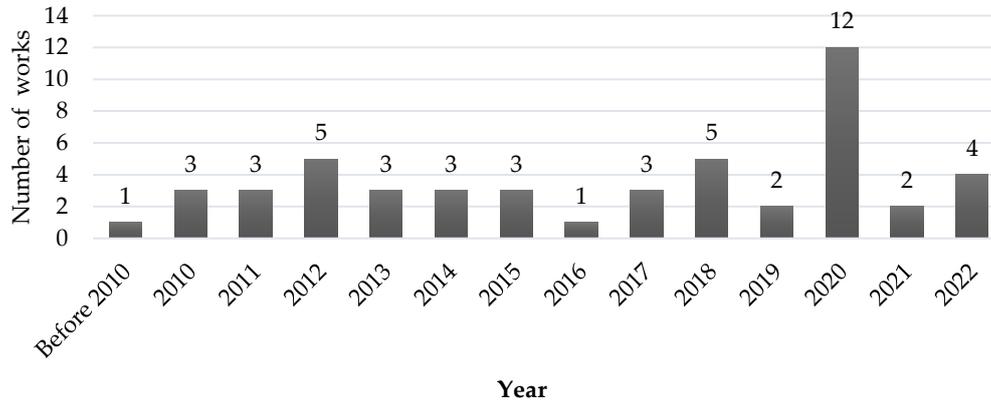

**Figure 3.** Studies by publication year.

*3.2. Catalogue Summary Statistics: By Datasets (n = 40)*

3.2.1. Types of Providers

We catalogued the RS datasets used in the curated publications, documenting their sources or providers. We categorized the sources into three groups: Government Agencies, Commercial Companies, and Academic Institutions (Table 2). The majority of the datasets (n = 28) were obtained from Government Agencies worldwide. Readers can refer to [10] for detailed information about the curated RS datasets and their corresponding providers and publications that featured them.

**Table 2**. Dataset sources/providers by category (government, commercial, academic) and provider's region.

| Provider Category | Provider Region | Providers | Number of Datasets |
|---|---|---|---|
| Government Agency | Asia | Earth Remote Sensing Data Analysis Center of Japan | 1 |
| | | National Catalogue Service For Geographic Information of China | 1 |
| | | National Geomatics Center of China | 1 |
| | Europe | European Space Agency (ESA) | 5, including 1 dataset co-provided with Université Catholique de Louvain |
| | | German Aerospace Center | 1 |
| | | Ordnance Survey (UK) | 3 |
| | | UK Centre for Ecology & Hydrology (UK CEH) | 2 |
| | America | Multi-Resolution Land Characteristics Consortium (USA) | 1 |
| | | National Aeronautics and Space Administration (NASA) | 6, including one dataset co-provided with NOAA |
| | | National Centers for Environmental Information, NESDIS, NOAA, U.S. Department of Commerce | 2, including one dataset co-provided with NASA |
| | | National Oceanic and Atmospheric Administration National Geophysical Data Center (USA) | 1 |
| | | Panama's Ministry of the Environment (MiAmbiente; formerly ANAM) | 1 |
| | | U.S. Geological Survey | 2 |
| | | U.S. National Geospatial-Intelligence Agency (NGA) | 1 |
| | | USDA's Farm Service Agency (FSA) | 1 |
| Commercial Company | America | Environmental Systems Research Institute (ESRI) | 2 |
| | | Google | 2 |
| | | Maxar Technologies | 1 |
| | | Microsoft | 1 |
| | Europe | OpenStreetMap (Foundation) | 1 |
| | | Vision On Technology | 1 |
| Academic Institute | Asia | Data Center for Resources and Environmental Sciences of the Chinese Academy of Science | 1 |
| | | Star Cloud Data Service Platform, Pengcheng Laboratory | 1 |
| | Europe | Copernicus Open Access Hub | 1 |
| | | Université Catholique de Louvain | 1 dataset co-provided with European Space Agency (ESA) |
| | America | Natural Resource Spatial Analysis Laboratory, Institute of Ecology, University of Georgia | 1 |

| | | 40 (co-provided datasets are only counted once) |
|---|---|---|
| Total | | |

### 3.2.2. Geographic Coverage

Table 3 presents the geographic coverage of the 40 catalogued datasets in our proof-of-concept catalogue. A considerable portion (about 60%) of these datasets encompasses a global scale, while the majority of the remaining datasets are available at a national or regional level. There are more datasets available for the USA and the UK compared with other countries in our proof-of-concept catalogue. Additionally, it is worth noting that we have included a dataset specifically covering oceanic regions. Readers can refer to [10] for detailed information about the curated datasets and their geographic coverage.

**Table 3.** Dataset coverage.

| Dataset Region | Covered Areas | Number of Datasets |
|---|---|---|
| Global | Global | 24 |
| | Global Coastal Zone (Ocean) | 1 |
| Americas | North America | 1 |
| | Panama | 2 |
| | USA | 4 |
| Asia | China | 2 |
| Africa | Kenya | 1 |
| Europe | England | 1 |
| | UK | 4 |
| Total | | 40 |

### 3.2.3. Costs

We also examined (and recorded in a dedicated metadata field in our catalogue) the cost of acquiring each of the curated RS datasets. Out of the 40 RS datasets in our proof-of-concept catalogue, 33 are freely accessible (free of charge), while the remaining seven require a purchase in order to use them for research purposes. Readers can refer to [10] for detailed information about the curated datasets and their cost status.

### 3.2.4. First Available Year and Update Frequency

Of the 40 datasets included in our proof-of-concept catalogue, 14 datasets were initially made available before the year 2000. A further 17 datasets were first released between 2000-2010, while the 9 remaining datasets became available for the first time after 2010. At least 21 out of the 40 curated datasets (52.5%) are still being updated as of 2022/2023 at various (usually regular) intervals since their initial release. The recorded update frequencies for these datasets range from daily updates and updates every ten days to monthly and quarterly updates to updates every five to ten years. Readers can refer to [10] for detailed information about the curated RS datasets and their corresponding first available year and update frequency details (where applicable and available).

3.2.5. Resolutions

Resolution is one of the critical factors in the selection of RS datasets as it needs to match the corresponding requirements of the application at hand. Table 5 provides details of the various dataset resolutions in our proof-of-concept catalogue. The most common dataset resolutions in the catalogue are 1km (7 datasets) and 30m (a further 7 datasets). Commercial companies such as Google Maps and Bing Maps provide RS datasets with notable resolution advantages (15cm–15m). It is important to note that some datasets offer multiple resolutions, as they contain remote sensing images captured in various bands. In such cases, panchromatic images generally have higher resolutions compared to multi-spectral images. Readers can refer to [10] for detailed information about the curated RS datasets and their various resolutions.

**Table 5.** Dataset resolutions (where available and recorded in our proof-of-concept catalogue). Some datasets offer multiple resolutions and therefore had to be counted more than once in this table.

| Resolution Type | Resolution | Number of Datasets |
|---|---|---|
| Length | >10km | 1 |
| | 1km | 7 |
| | 800m | 1 |
| | 300m | 1 |
| | 250m | 2 |
| | 84m | 1 |
| | 30m | 7 |
| | 20m | 4 |
| | 10m | 2 |
| | 5m | 1 |
| | 2.4m / Multispectral | 2 |
| | 0.6m / Panchromatic | 3, including the same two multispectral datasets counted in above row, as they are also panchromatic |
| | 0.15–0.5m | 4 |
| | N/A | 2 |
| Scale | 1:10,000 | 1 |
| | 1:100,000 | 1 |
| | 1:250,000 | 1 |
| | 1:7,000,000 | 1 |

*3.3. GeoExposomics.org Online Portal and Searchable Catalogue*

The project's portal offers a searchable version of the catalogue, and can be accessed at [11] (Figures 2 and 3). It allows users to search and filter dataset records by health conditions/application areas, cost, covered area(s), provider, or a combination of one or more of these elements. Users can also filter the associated research article records by health conditions/application areas and/or dataset name.

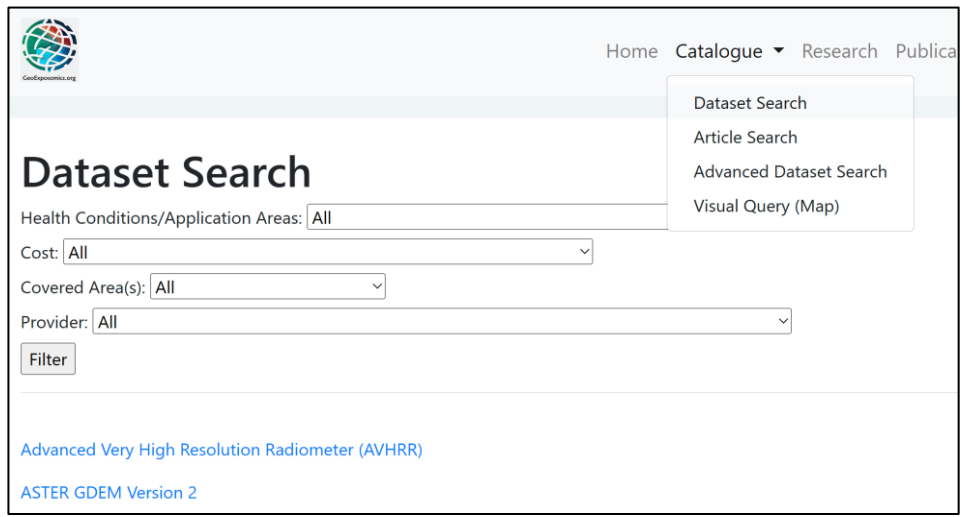

**Figure 2.** An online searchable version of the catalogue can be accessed by clicking the 'Catalogue' menu on the project portal's navigation bar. Various search modes and result filtering options are currently available.

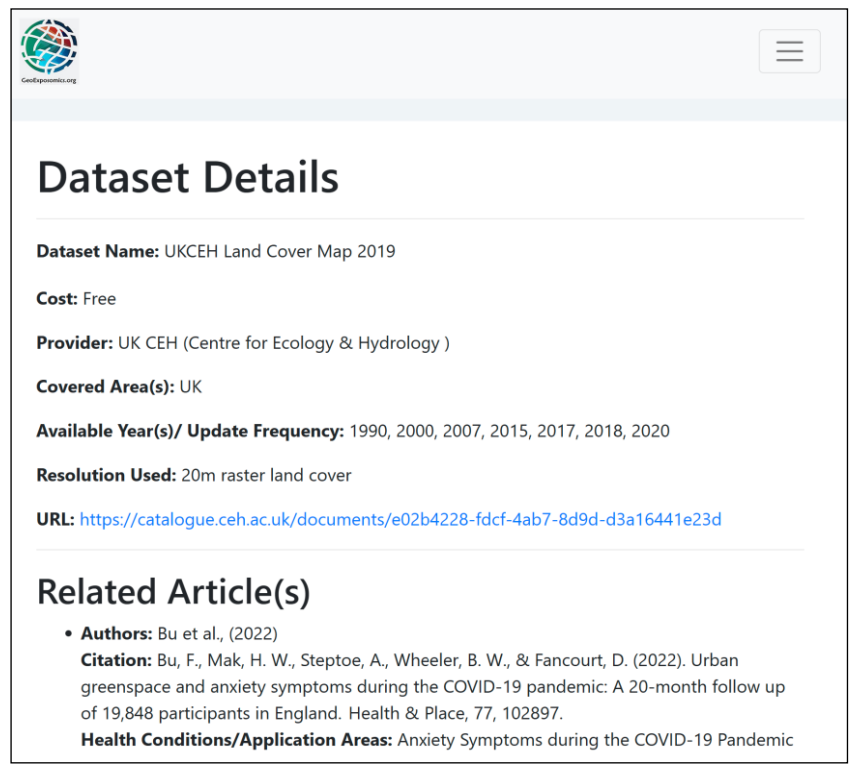

**Figure 3.** A typical metadata record in the online catalogue. This screenshot shows the mobile version of the interface with the 'hamburger menu' (top right) for site navigation.

*3.4. Evaluation Feedback and Second Iteration Improvements*

The evaluation feedback received from our panel of five stakeholder representatives about the online portal was positive and provided a couple of valuable suggestions that were later implemented in a second iteration of the interface (released in late June 2023), namely a suggestion "to implement free text search and ensure it can tolerate different spelling variants (e.g., British and American) and misspellings of health applications and place names" and another one to have a handy "online map to help locating datasets". In response, we implemented a free text search option of datasets (by health applications or covered areas) using a custom Soundex algorithm [12] (Figure 4) and introduced a visual query option using a map with clickable hotspots (Figure 5). It should be noted that while free text search using Soundex is an improvement, it still cannot help when a query string is a synonym, i.e., a very different word or phrase not in the catalogue as such; for example, when searching for 'coronavirus disease' instead of 'COVID-19'. This remains a limitation of the current search interface.

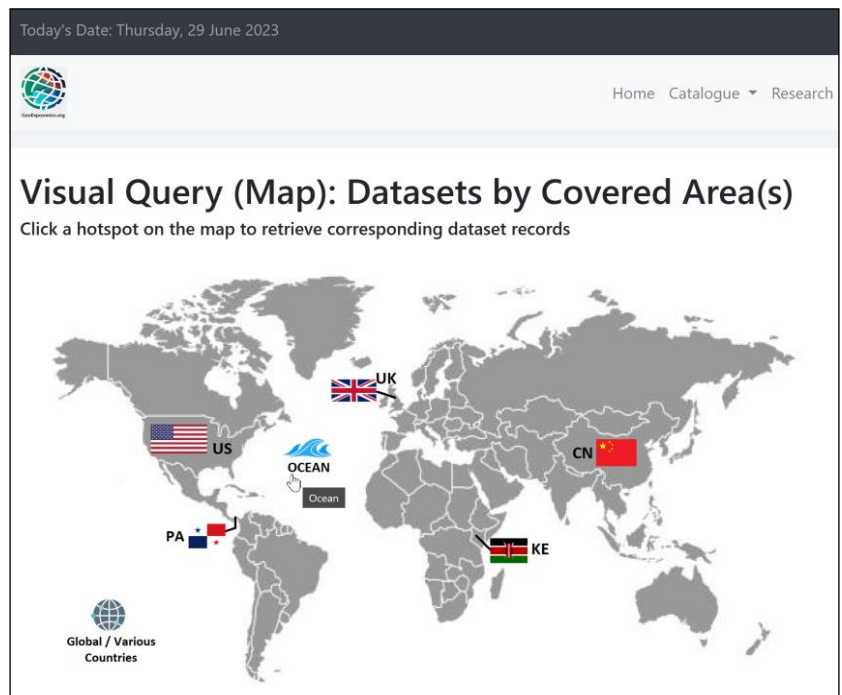

**Figure 4.** Following user evaluation feedback, we implemented a free text search option of datasets (by health applications or covered areas) using a custom Soundex algorithm. As an example of how this algorithm works, searching health conditions for 'haemorrhagic fever' (British spelling), 'hemorrhagic fever' (American spelling), or 'hemoragic fever' (typo/misspelling, as in the above screenshot) will all retrieve the correct/same results.

**Figure 5.** User evaluation feedback included a suggestion to provide an online map to visually locate datasets. In response, we introduced an option allowing users to click location 'hotspots' on a map of the world with country flags and other relevant icons to retrieve corresponding dataset records.

*3.5. Dissemination*

The portal was widely advertised on social media upon its initial launch in June 2023 and subsequent updates during June and July 2023 using the P.I.'s (MN Kamel Boulos) accounts on Facebook, Instagram, LinkedIn, and Twitter (>2700 combined followers across the four platforms), with additional reposts via other relevant public LinkedIn and Twitter accounts. Between 15 June and 31 July 2023, the portal received >550 unique visitors from more than 25 countries according to anonymous server log statistics generated by AWStats tool, an open-source server log file analyzer for advanced statistics, using site visitors' IP geolocation. We also have a companion Facebook page where relevant news items and updates are regularly shared with a growing community [13].

**4. Discussion**

RS techniques can help us better map and investigate the exposomic determinants of disease (population-level exposures) by providing spatially and temporally resolved data over time at multiple geographic scales. For example, satellite imagery can be used to monitor air quality and temperature, and detect the presence of pollutants or heat waves and extreme weather events that may have adverse effects on human health [14,15,16]. RS data can also aid in identifying patterns of land use and land cover that contribute to the spread of diseases such as malaria [17] and dengue fever [18]. However, discovering and locating relevant RS datasets for use in a particular health study can often prove challenging, hence the need for metadata catalogue services to assist researchers in this task.

*4.1. Metadata Catalogue Services*

Catalogue services, such as our proof-of-concept portal [11], allow for the publishing and searching of collections of descriptive information (or metadata) about resources: data, services, and related information objects, or in our case, metadata about RS datasets of interest. As such, metadata catalogue services are essential for supporting "the discovery and binding to registered information resources within an information community" [19], which in the case of our project would be the 'RS and population health and medicine research communities'. It is worth noting here that the European Union's INSPIRE (Infrastructure for Spatial Information in Europe) portal in partnership with UNIGIS offer a highly relevant online training module entitled 'metadata and catalogue services' and aimed at data users and data providers [20].

*4.2. Reflections on Our Catalogue Summary Statistics*

In our analysis of the publications included in our proof-of-concept catalogue, we observed that more studies were published in geography and environmental science journals than in public health journals. However, with our very small convenience sample of five experts and 50 curated publications, one cannot boldly generalize this observation. Nevertheless, there are a couple of reasons as to why RS data are still relatively underused in population health studies.

Firstly, some researchers with a primary background in health might be unfamiliar with RS data and their potential applications in health studies. Secondly, in order to use RS data proficiently, researchers should also be familiar with, or seek interdisciplinary collaborations with geomatics researchers who are familiar with, the different types and formats of RS data. In general, RS data come from imagery. To digitally manage and process imagery data, RS data are constructed in the form of pixels (cells) or in raster, the rectangular arrays of numbers containing both geometric and radiometric characteristics from a surveyed area, often extended with other attribute data. For efficient data management and analysis, raster data are written and managed in several different file formats to encode a variety of geographic data structures. Therefore, conceptually understanding and utilizing RS data in raster format may be challenging for health researchers who are usually more familiar with numbers-only data written in a simpler structure.

Another observation relates to the study areas of the catalogued publications and their corresponding dataset coverage. Only a handful of studies in our sample examined problems in the Global South. Again, this observation cannot be readily generalized given our small sample size, but it still has some merit and is worth considering. Unarguably, there are many public health and environmental challenges to address in the Global North, especially in light of climate change [6] and population ageing. However, more research should be conducted to investigate such challenges in the Global South, considering (and addressing) the uneven development in research experience, capacity, and infrastructure there.

*4.2.1. Scale and Resolution Matter*

Thanks to rapid technological developments, it is becoming increasingly easier nowadays to use RS data with low, medium, and high spatial resolution. A commercial satellite offers a spatial resolution of 30cm x 30cm, which is about the finest spatial resolution for now. It is not always the case, but a higher spatial resolution could be helpful in a closer examination of the impact of the environment on population health as it can provide finer details. However, obtaining higher resolution RS data is costly, and there is not always a clear consensus about what level of resolution would be optimal for use in a given health application or study. Considering the temporal resolution, i.e., the frequency of the revisit period, could be another challenge when controlling for the continuing interaction between humans and the environment.

*4.2.2. The Importance of Data Linkages*

The majority of our catalogued RS datasets are publicly available free of charge. Two of the most notable and frequently cited RS data portals in our proof-of-concept catalogue are NASA's Earthdata [21] and ESA's Sentinel Online [22]. These data portals provide researchers and any interested members of the public with one-stop services covering a range of user needs and topics from RS data access to various research themes and case studies to self-learning material. However, these data portals only serve as a platform for RS data and related information, without offering any linkages to population-related data. In contrast, the applications of GIS (Geographic Information Systems) have become standard in the public health literature, with the US Centers for Disease Control and Prevention (CDC) now operating a database service linking relevant demographic, social, and environmental data, as well as GIS software and tools [23]. Future development in RS data portals should explore potential ways of linking RS data to other datasets to enable comprehensive research in disciplines requiring such linkages, such as population health and medicine.

*4.3. Future Directions: Turning Current Limitations into Opportunities for Future Improvements*

Our current small sample size and online user interface limitations (e.g., free text search not catering for synonyms) can be used to inform, and set the desiderata for, possible future developments. A follow-on project would conduct a much wider stakeholder consultation and comprehensively catalogue more datasets covering more countries/locations and more health conditions/application areas. Moreover, an online form could be made available on the portal for visitors to propose and contribute metadata details of new records to be added to the Catalogue (if deemed suitable following moderation to ensure quality and completeness).

As the catalogue expands, revisiting and improving our metadata schema, elements and descriptions' consistency and organization, expanding them as appropriate, and applying the relevant standards of metadata and catalogue services [19,20,24] will become imperative. A protocol with the necessary mechanisms for catalogue ongoing maintenance will need to be established to ensure the continued currency of the curated metadata records and the details stored within.

We could also look into providing a ChatGPT-like [25-28], natural language-based AI (artificial intelligence)-driven conversational user interface (UI) for querying the catalogue, whereby users use free text (plus, optionally, a more advanced interactive map to select desired locations) to describe in their own way and words a research question or health application they are currently investigating in a particular locale in order to retrieve all relevant records from the catalogue. This would enhance discovery and explorability by allowing the search UI to more flexibly adapt to users' needs in more natural ways instead of asking users to adapt to and master a rather rigid (preprogrammed) and less forgiving interface.

The implementation of a ChatGPT-like interface will make use of Retrieval-Augmented Generation (RAG) whereby our metadata catalogue becomes an external knowledgebase to an existing Large Language Model (LLM), such as GPT-4, in order to provide more relevant and better cross-linked search results in response to user's prompts [29] (Figure 6).

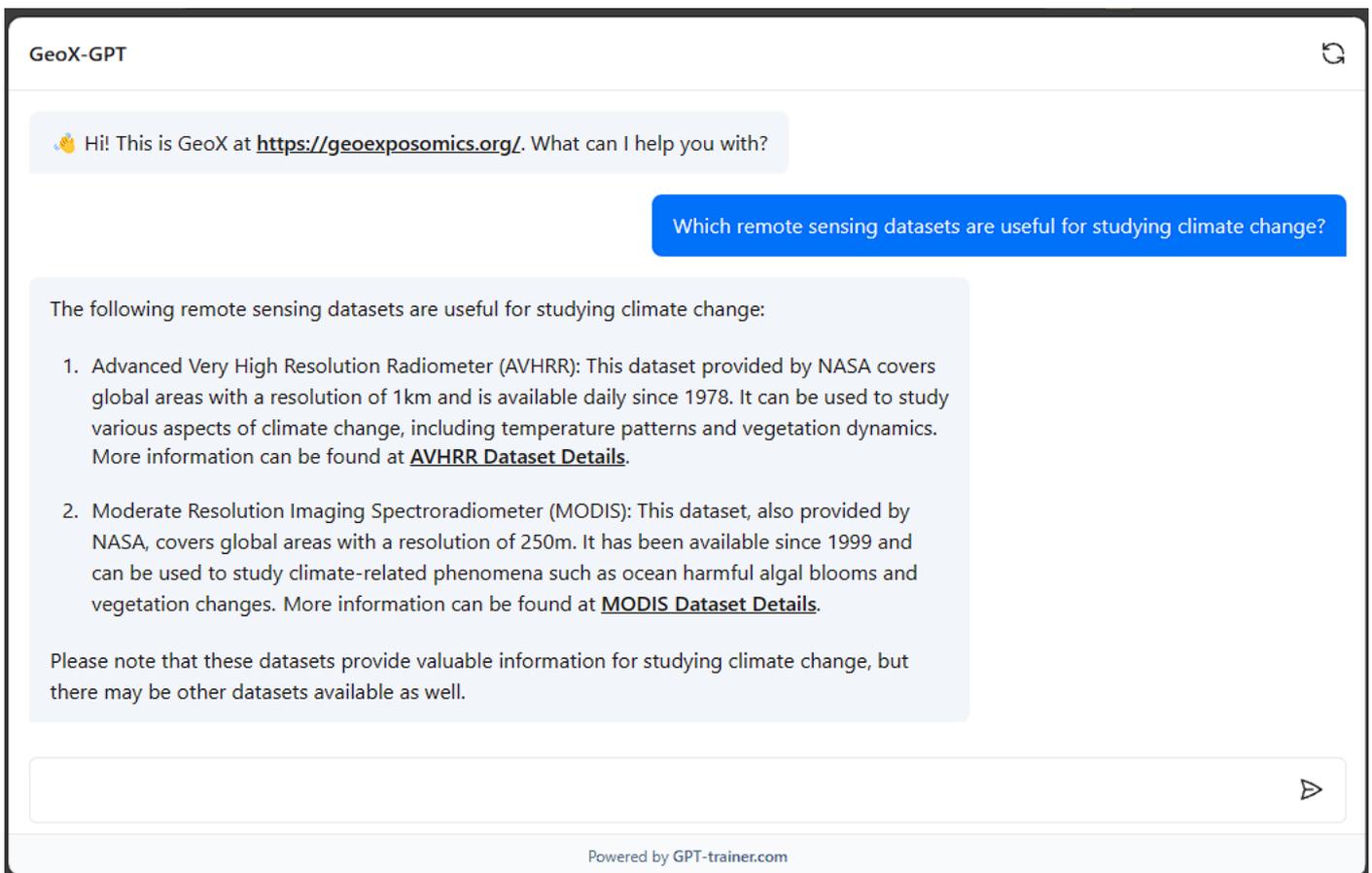

**Figure 6.** Screenshot of GeoX-GPT, an early preview of a possible future ChatGPT-like interface for our metadata catalogue. This experimental RAG implementation was developed in GPT-trainer [30].

## 5. Conclusions

Earth observation data play an important role in understanding the relationship between environmental factors and human health, offering valuable insights into the exposomic determinants of disease (population-level exposures). During 2023, the GeoExposomics Project Team built and published a proof-of-concept online RS data catalogue to further facilitate human health research in exposomics, through an international online collaboration between the Project Team members themselves (who are based in five countries spanning four continents) and with an external panel of five leading scholars from three countries.

In this article, we conducted an analysis of the content of our proof-of-concept catalogue and described the steps taken to convert it into a user-friendly online searchable format over two development iterations informed by user feedback. We also provided some reflections on our catalogue summary statistics, our current project limitations, and possible future developments. It is hoped future, more comprehensive versions of this service will offer a complete one-stop shop for discovering health-related RS datasets from multiple providers worldwide, enabling more interdisciplinary researchers and human exposomics studies to discover and use Earth observation data in combination with other relevant data to reveal fresh insights that could improve our understanding of relevant diseases, and hence contribute to the development of better-optimized prevention and management strategies to tackle them.

**Author Contributions:** Conceptualization, M.N.K.B.; methodology, M.N.K.B. (Main) and K.K.; validation, H.Z. and M.N.K.B.; formal analysis and investigation, K.K., G.Z. and M.N.K.B.; resources, M.N.K.B.; data curation, K.K. (Main), G.Z. (Main) and M.N.K.B.; writing—original draft preparation, K.K., G.Z. and M.N.K.B.; writing—review & editing, K.K., G.Z., M.N.K.B., H.Z., M.V.I, B.B. and A.D.; visualization, K.K., G.Z. and M.N.K.B.; supervision and project administration, M.N.K.B. (Main/Lead), K.K., M.V.I, B.B. and A.D.; funding acquisition, M.N.K.B. (P.I.) with M.V.I, B.B. and A.D. (Co-Is). All authors have read and agreed to the published version of the manuscript.

**Funding:** This research was funded by the International Society for Photogrammetry and Remote Sensing (ISPRS) under ISPRS Scientific Initiatives 2023 (SI2023 Awards).

**Data Availability Statement:** The proof-of-concept catalogue data covered in this paper are publicly and freely available from doi: 10.5281/zenodo.8186444 and https://geoexposomics.org/

**Acknowledgments:** The authors would like to express their gratitude to the panel of stakeholder representatives that were involved in the project.

**Conflicts of Interest:** The authors declare no conflict of interest.